\documentstyle[12pt,twoside,fleqn,espcrc1,graphicx,]{article}
\newcommand\subtitle{
\vspace{-57mm} \vspace{-1\baselineskip}
\noindent \mbox{\scriptsize \it
To be published in Proc.\ COSPAR Coll.\ ``The Outer Heliosphere: The Next Frontiers'', eds.\
H.J.~Fahr et al.
}\vspace{55mm}}

\newcommand{\aindex}{$^{\tiny \rm a}$}

\title{Antiprotons below 200 MeV in the interstellar medium:\\
perspectives for observing exotic matter signatures}

\author{I.V. Moskalenko\aindex%
\thanks{NRC Senior Research Associate}%
\thanks{also Institute of Nuclear Physics, M.V.Lomonosov Moscow State University, 
Moscow 119899, Russia},
E.R. Christian\aindex, 
A.A. Moiseev\aindex, 
J.F. Ormes\address{NASA/Goddard Space Flight Center, Greenbelt, MD 20771, USA},
and
A.W. Strong\address{Max-Planck-Institut f\"ur extraterrestrische Physik, 
85741 Garching, Germany}}

\begin{document}

\maketitle
\subtitle

\begin{abstract}
Most cosmic ray antiprotons observed near the Earth are secondaries produced
in collisions of energetic cosmic ray (CR) particles with interstellar gas.
The spectrum of secondary antiprotons is expected to peak at $\sim2$ GeV and
decrease sharply at lower energies.  This leaves a low energy window in
which to look for signatures of exotic processes such as evaporation of
primordial black holes or dark matter annihilation.  In the inner
heliosphere, however, modulation of CRs by the solar wind makes
analysis difficult. Detecting these antiprotons outside the heliosphere
on an interstellar probe removes most of the complications of modulation.
We present a new calculation of the expected secondary antiproton flux 
(the background) as well as
a preliminary design of a light-weight, low-power instrument for the interstellar
probe to make such measurements.
\end{abstract}

\section{INTRODUCTION}
The nature and properties of the dark matter that may constitute up to 70\%
of the mass of the Universe has puzzled scientists for decades, e.g.\ see
\cite{Trim87}. It may be in the form of weakly interacting massive particles
(WIMPs), cold baryonic matter, or primordial black holes (PBHs). 
It is widely believed that the dark matter may manifest itself 
through annihilation (WIMPs)
or evaporation (PBHs) into well-known stable particles. The problem, however,
arises from the fact that a weak signal should be discriminated from an
enormous cosmic background, including a flux of all known nuclei, electrons, 
and $\gamma$-rays.
Antiproton measurements in the interstellar space could provide
an opportunity to detect a signature of such dark matter 
(see \cite{Well99} and references therein).

High energy collisions of CR particles with interstellar gas
are believed to be the mechanism producing the majority of CR 
antiprotons. Due to the kinematics of the process they are created 
with a nonzero momentum providing a low-energy ``window'' where exotic
signals can be found. It is therefore important to know accurately
the background flux of interstellar secondary antiprotons
and to make such measurements outside the heliosphere to avoid 
any uncertainties due to solar modulation.

\begin{figure}[tb]
\begin{minipage}[t]{160mm}
\includegraphics[width=79mm]{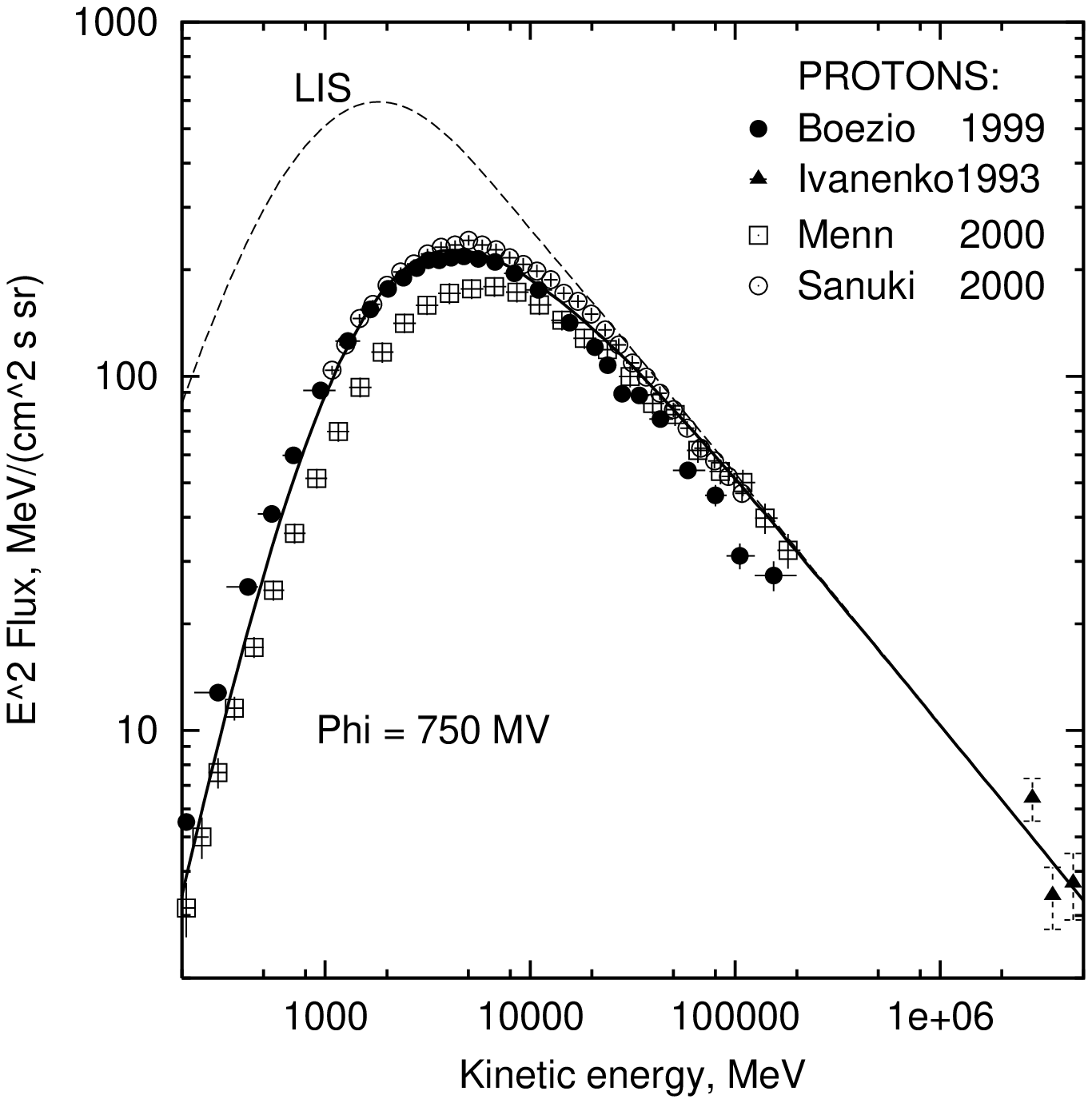}
\hspace{\fill}
\includegraphics[width=79mm]{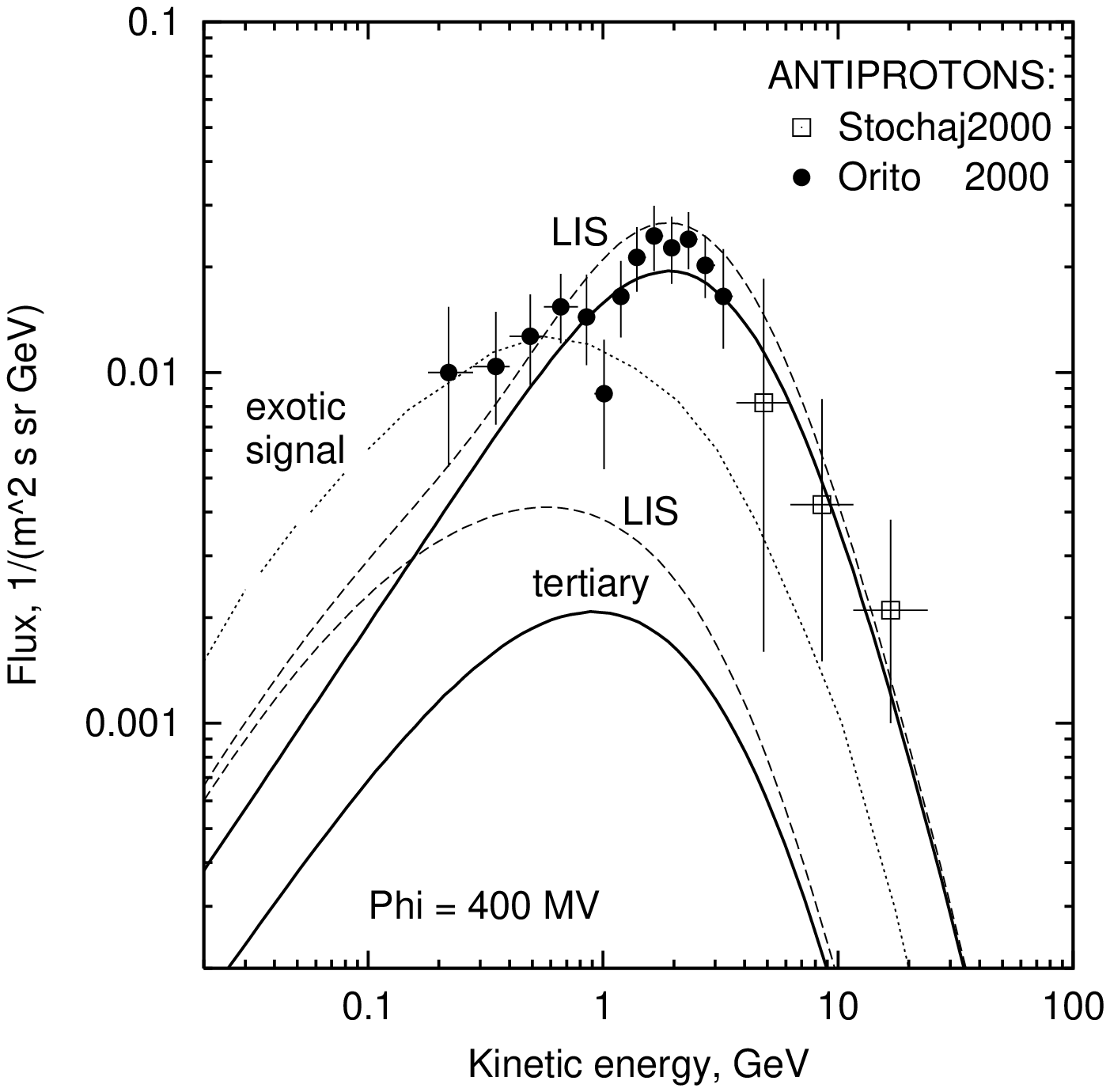}
\caption{{\it Left:} Calculated proton interstellar spectrum (LIS) and 
modulated spectrum ($\Phi=750$ MV) with data \cite{Menn00,Boez00,Sanu00,Sokol}.
{\it Right:} Calculated $\bar{p}$ interstellar spectrum (LIS) and 
modulated spectrum ($\Phi=400$ MV) with data \cite{Orit00,Stoc00}.
The top curves are the total background. 
The ``tertiary'' components (LIS and modulated) are shown separately. 
Also shown is an example exotic signal \cite{Berg99} extended to lower 
energies.}
\label{fig1}
\end{minipage}
\end{figure}

\section{CALCULATIONS OF THE ANTIPROTON BACKGROUND}

We made a new calculation of the CR $\bar{p}$ flux in our
model (GALPROP) which aims to reproduce observational data of many
different kinds: direct measurements of nuclei, $\bar{p}$'s, $e^\pm$'s,
$\gamma$-rays and synchrotron radiation \cite{MSR98,SM98,SMR00}. 
The model was significantly improved, and entirely rewritten in C++. 
The improvements involve various optimizations relative to our older
FORTRAN version, plus treatment of full reaction networks, an extensive
cross-section database and associated fitting functions, and
the optional extension to propagation on a full 3D grid.
For this calculation, we used
a cylindrically symmetrical geometry with parameters that have been 
tuned to reproduce observational data \cite{SMR00}.

The propagation parameters including diffusive reacceleration
have been fixed using boron/carbon and $^{10}$Be/$^9$Be ratios.
The injection spectrum was chosen to reproduce local CR measurements, 
$\sim\beta\rho^{-2.38}$, where $\beta$ is the particle
speed, and $\rho$ is the rigidity. The parameters used: the diffusion coefficient, 
$D_{xx}=4.6\times10^{28}\beta(\rho/3\,{\rm GV})^{1/3}$ cm$^2$ s$^{-1}$, 
Alfven speed, $v_A=24$ km s$^{-1}$, and the halo size, $z_h=4$ kpc.

We calculate $\bar{p}$ production and propagation using the basic formalism
described in \cite{MSR98}. To this we have added $\bar{p}$ annihilation 
and treated inelastically scattered $\bar{p}$'s as a separate 
``tertiary'' component (see \cite{TanNg} for the cross sections). 
The $\bar{p}$ production by nuclei with $Z\geq2$ is calculated in two ways:
employing scaling factors \cite{MSR98}, and using effective
factors given by Simon et al. \cite{Simo98}, who make use of the DTUNUC code, 
and which appears to be more accurate than simple scaling.
(The use of Simon et al. factors is consistent since their adopted
proton spectrum resembles our spectrum above the $\bar{p}$ production threshold.) 
The effect on the $\bar{p}$ flux at low energies is
however small, and the two approaches differ by about 15\%. 

We believe our calculation is the most accurate so far since we
used a self-consistent propagation model and the most accurate production
cross sections \cite{Simo98}. 
The results are shown in Figure \ref{fig1}. 
The upper curves are the local interstellar flux (LIS) and the lower are modulated
using the force-field approximation. 
The two lowest curves in Figure \ref{fig1} (right) show separately the contribution of 
``tertiary'' $\bar{p}$'s, which is the dominant component at low energies. 
The adopted nucleon injection spectrum, after propagation,
matches the local one. There remains some excess of $\bar{p}$'s.
The excess for the lowest energy points is at the 1 $\sigma$ level.

Many new and accurate data on CR nuclei, diffuse gamma rays,
and Galactic structure have appeared in the last decade; this allows
us to constrain propagation parameters so that the limiting
factor now becomes the isotopic and particle production cross
sections.
%We are approaching the limit where to make new progress in such calculations
%we need a new, more accurate, measurement of the $\bar{p}$ production cross section.
At this point we cannot see how to increase the predicted intensity
unless we adopt a harder nucleon spectrum at the source in contradiction
with constraints from high energy $\bar{p}$ data \cite{MSR98,SMR00}.
More details will be given in a subsequent paper.

\section{ANTIPROTON DETECTOR}

\begin{figure}[tb]
\begin{minipage}[t]{160mm}
\includegraphics[width=56mm]{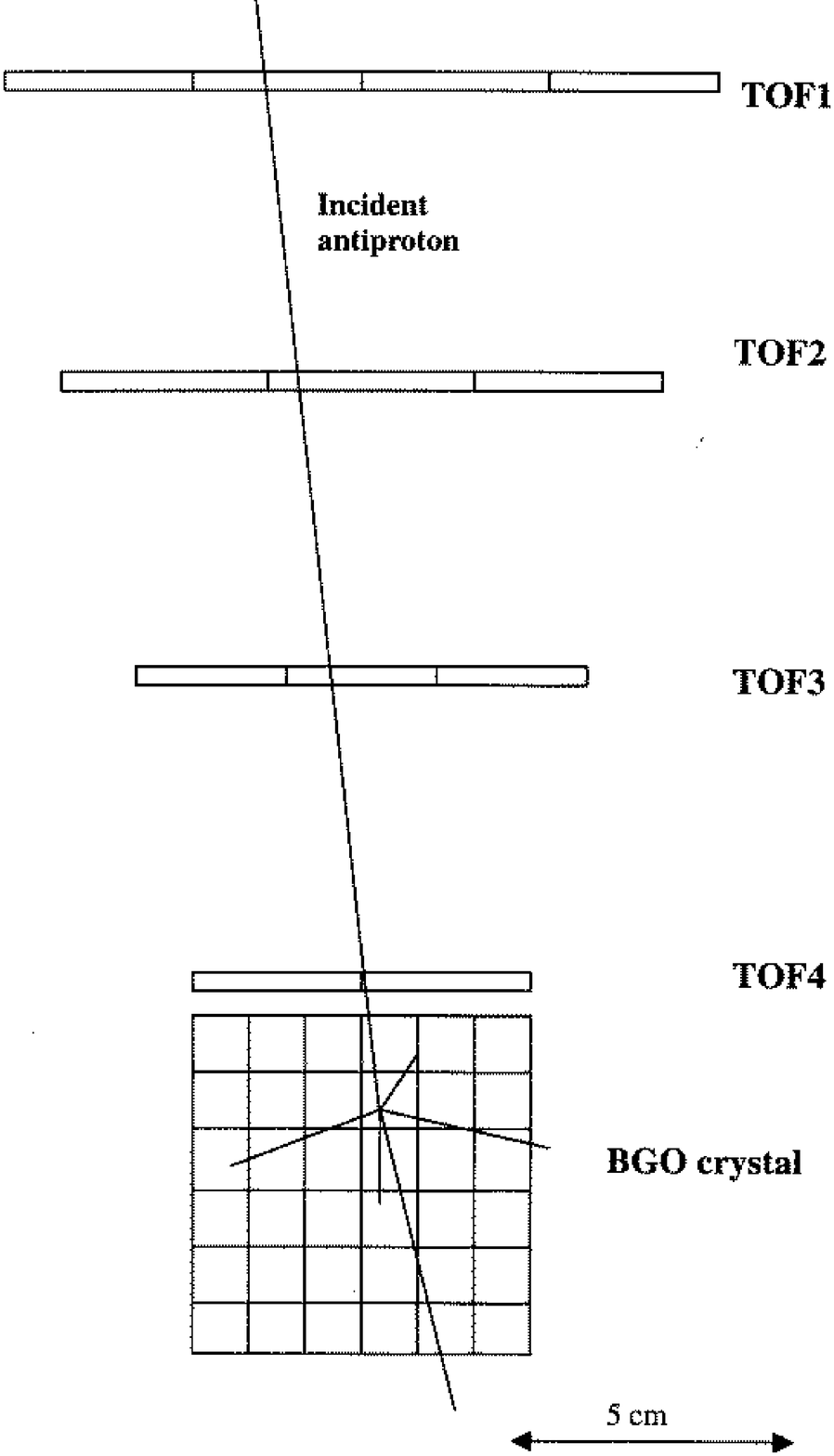}
\hfill
\includegraphics[width=90mm]{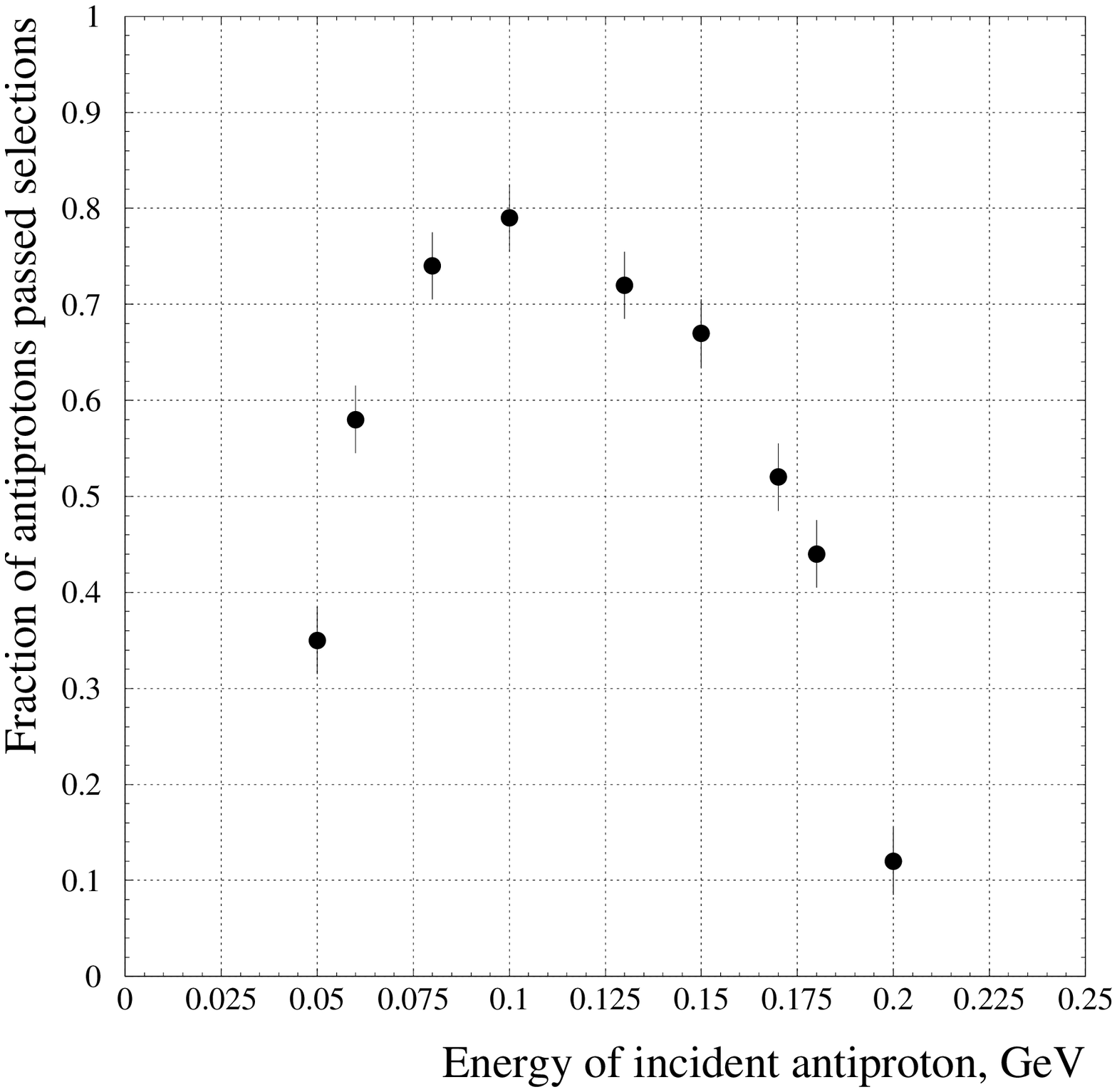}
\caption{{\it Left:} Preliminary design of the detector. 
{\it Right:} efficiency for antiproton detection after selections applied.}
\label{fig2}
\end{minipage}
\end{figure}

Very limited weight and power will be available for any experiment on board
an interstellar probe \cite{probe}. We therefore propose a simple instrument which
is designed to satisfy these strict constraints \cite{Well99}.
We base our design (Figure \ref{fig2} left) on a cube of heavy 
scintillator (bismuth germanium oxide
[BGO]), with mass of the order of 1.5 kg. The cube, 42 g cm$^{-2}$ thick,
will stop antiprotons and protons of energy below 250 MeV. A time-of-flight
(TOF) system is used to select low-energy particles. The particles with energy
less than $\sim50$ MeV will not penetrate to the BGO crystal through the TOF
counters, setting the low-energy limit.

The separation of antiprotons from protons is the most challenging aspect
of the design. A low-energy proton (below 250 MeV) that would pass the
TOF selections cannot deposit more than its total kinetic energy in the 
block. Therefore an event will be required to deposit $>300$
MeV to be considered an antiproton. A proton can deposit comparable energy 
in this amount of material only through hadronic interaction, which
our Monte Carlo simulations show 
requires a proton with energy $>500$ MeV. The TOF system can effectively 
separate low-energy particles ($<250$ MeV) from such protons 
and heavier nuclei.

As a conservative estimate, we assume that all protons with energy
$>500$ MeV have the potential to create a background of ``$\bar{p}$-like''
events and their integral flux in interstellar space is 
somewhat uncertain but would be 
$\sim1$ cm$^{-2}$ s$^{-1}$ sr$^{-1}$. The exotic $\bar{p}$ signal, to be
significantly detected above the background, should be of the order of $10^{-6}$
cm$^{-2}$ s$^{-1}$ sr$^{-1}$ in the energy interval 50--200 MeV, which
corresponds to an expected signal/$\bar{p}$-background ratio of $\sim 10$. We thus
can allow only one false antiproton in $10^7$ protons. 

Simulations to date indicate that the current design will have rejection
power of $\sim 2\times10^6$. We expect to get the next factor of five
by fine-tuning the design and selections. The efficiency of antiproton 
selection is shown in Figure \ref{fig2} (right). 
The antiproton rate will be 0.1--1 particle per day, and
the statistical accuracy will be $\sim10$\% after 3 years of observation.

I.V.M. acknowledges support from NAS/NRC Senior Research Associateship Program.

\end{document}